\documentclass[utf8]{frontiersSCNS} 

\usepackage{url,lineno,microtype}
\usepackage[onehalfspacing]{setspace}
\usepackage{comment}

\def\keyFont{\fontsize{8}{11}\helveticabold }
\def\firstAuthorLast{Naddaf {et~al.}} 
\def\Authors{Mohammad-Hassan Naddaf\,$^{1,2,*}$, Bo\.zena Czerny\,$^{1}$, Ryszard Szczerba\,$^{3}$}
\def\Address{$^{1}$Center for Theoretical Physics, Warsaw, Poland \\
$^{2}$Nicolaus Copernicus Astronomical Center, Warsaw, Poland \\
$^{3}$Nicolaus Copernicus Astronomical Center, Torun, Poland  }
\def\corrAuthor{Mohammad-Hassan Naddaf}

\def\corrEmail{naddaf@cft.edu.pl}

\begin{document}
\onecolumn
\firstpage{1}

\title[BLR size in FRADO]
{BLR size in Realistic FRADO Model: The role of shielding effect} 

\author[\firstAuthorLast ]{\Authors} 
\address{} 
\correspondance{} 

\extraAuth{}

\maketitle

\begin{abstract}
The effective size of Broad Line Region (BLR), so-called the BLR radius, in galaxies with active galactic nuclei (AGN) scales with the source luminosity. Therefore by determining this location either observationally through reverberation mapping or theoretically, one can use AGNs as an interesting laboratory to test cosmological models. In this article we focus on the theoretical side of BLR based on the Failed Radiatively Accelerated Dusty Outflow (FRADO) model. By simulating the dynamics of matter in BLR through a realistic model of radiation of accretion disk (AD) including the shielding effect, as well as incorporating the proper values of dust opacities, we investigate how the radial extension and geometrical height of the BLR depends on the Eddington ratio [and blackhole mass], and modeling of shielding effect. We show that assuming a range of Eddington ratios and shielding we are able to explain the measured time-delays in a sample of reverberation-measured AGNs.
\tiny
 \keyFont{ \section{Keywords:} Active Galaxies, Broad Line Region, FRADO Model, Shielding Effect, Dust Opacity} 
\end{abstract}

\section{Introduction}

The delayed response of BLR to the variation of AGNs continuum is a probe to study the structure of BLR \citep{1998PASP..110..660P}. The main discovery is a linear $\log - \log$ scaling law between the size of BLR and the AGN monochromatic luminosity, so-called the radius-luminosity (RL) relation. It has been observationally discovered \citep{2000ApJ...533..631K,2004ApJ...613..682P,2009ApJ...705..199B, 2013ApJ...767..149B}, and theoretically explained by \citet{2011A&A...525L...8C} who proposed that failed dust-driven wind (FRADO) is responsible for the formation of BLR. However, the departures from the law seen for some sources with high Eddington ratios \citep{2015ApJ...806...22D, 2016ApJ...825..126D, 2018ApJ...856....6D, 2017ApJ...851...21G}, obviously requires the structure and extension of BLR to be studied in more detail, not only in observational side via reverberation-mapping (e.g. \citealp{2019FrASS...6...75P}), but also through theoretically-motivated models.
Provided that the RL relation is concretely justified for all already-observed sources  \citep{2019ApJ...883..170M, 2019ApJ...886...93R}, AGNs can be treated as astronomical candles to be used in cosmology \citep{2011ApJ...740L..49W, 2011A&A...535A..73H}.

In this article, we briefly report our pilot numerical study on the size of BLR based on the FRADO model. A concise background on the existing BLR models is addressed in section \ref{sec:BLRModels}. In section \ref{sec:FRADOModel} an introduction to FRADO model is provided, followed by section \ref{sec:Shielding} on shielding effect. After a short description of our numerical setup in section \ref{sec:Setup}, the method we adopted to find the BLR size is specified in section \ref{sec:BLRsize}. We discuss our results and conclude the article in the last two sections.

\section{BLR Models}\label{sec:BLRModels}

Broad Emission Lines are the most characteristic property of the AGN spectra known since many years (for a review, see \citealp{2015ARA&A..53..365N}). However the formation of BLR is not clear and still under debate. Different scenarios about the origin of the BLR material have been discussed during years. The proposed mechanisms can be divided into: Inflow Models \citep{2008ApJ...687...78H, 2017NatAs...1..775W}, Disk Instabilities \citep{1999A&A...344..433C, 2008A&A...477..419C, 2011ApJ...739....3W, 2012ApJ...746..137W}, or Disk Winds/Outflows. The latter class of models includes Magnetically-Driven winds \citep{1982MNRAS.199..883B, 1992ApJ...385..460E, 2014MNRAS.438.3340E}, Thermally-Driven winds \citep{1983ApJ...271...70B, 1989MNRAS.236..843C, 1997MNRAS.286..848W, 1999MNRAS.303L...1B}, and Radiatively-Driven winds consisting of Line-Driven winds \citep{1995ApJ...451..498M, 2010A&A...516A..89R}, and newly-born models of Dust-Driven winds.

The first model in the category of Dust-driven Winds is FRADO model \citep{2011A&A...525L...8C, 2015AdSpR..55.1806C, 2017ApJ...846..154C}. There is only another existing model in this category for the formation of BLR in which the static atmosphere of disk is puffed up by irradiation in the dust-dominated region \citep{2018MNRAS.474.1970B}. Differently from this static model, FRADO gives a dynamic view of the BLR. The advantage of the FRADO model is that the underlying  physics is known, the approximate compensation of the inflow/outflow leads to approximately symmetric lines, and the model is likely to produce also single component line instead of disky profiles if the vertical motion is strong, as expected at higher Eddington ratios. However, the model applies only to Low Ionization Lines, and for High Ionization Lines a line-driven outflow is likely favored.

\section{FRADO Model}\label{sec:FRADOModel}

FRADO model is one of the interesting theoretically-based models explaining the formation mechanism of the BLR in AGNs. 
It is based on the existence of dust at large radii where the AD atmosphere is cold enough \citep{2008MNRAS.383..581D}. The dusty material, i.e. clumps composed of dust and gas, is lifted up by the AD local radiation flux. Once reaching considerably high altitudes above the disk surface, dust content of clump gets evaporated by the strong radiation coming from the central parts of an AGN and then follows a subsequent ballistic motion and forms a failed wind.

In the basic 1-D analytical form of FRADO model, only the vertical component of local flux of accretion disk at a given radius was considered. It directly gives the inner/outer radius of BLR. Asymmetry in line profiles can be seen due to the dust evaporation; line shape parameter $\mathrm{FWHM}/ \sigma$ is consistent with data; and also the model predicts the dependence of virial factor on blackhole mass \citep{2015AdSpR..55.1806C, 2016ApJ...832...15C, 2017ApJ...846..154C}. However, the analytical local flux of Shakura-Sunyaev (S-S) disk \citep{1973A&A....24..337S} that is used in this model is constant with height; radiation from other radii and also the radial motion of material are neglected, as well as the inner boundary condition in S-S AD.

\begin{figure}
\begin{center}
\includegraphics[scale=0.6]{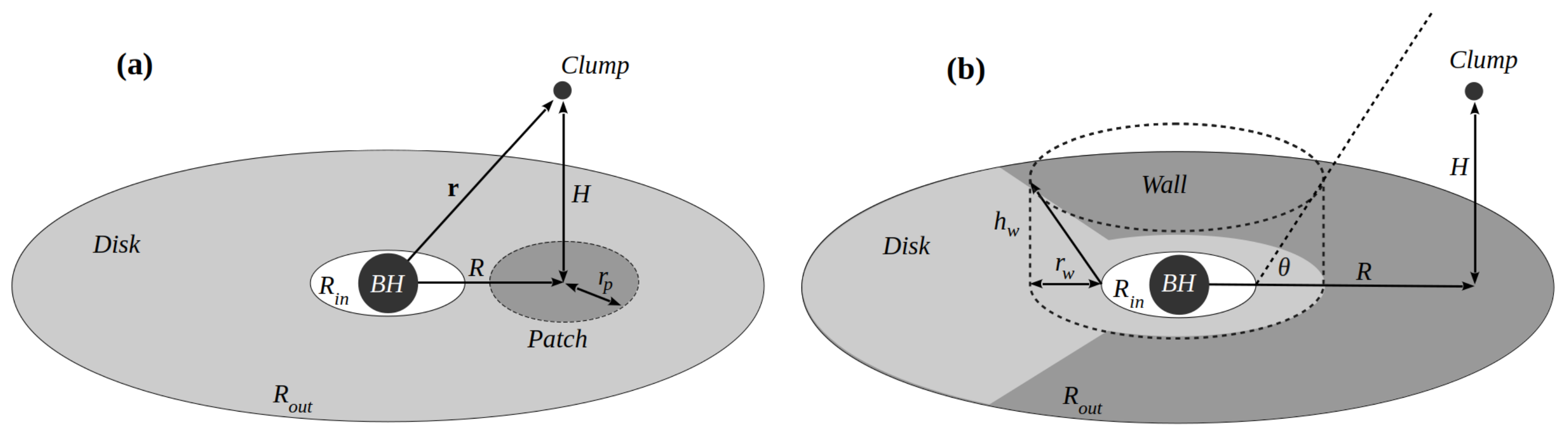}
\end{center}
\caption{Illustration of shielding models. The disk is shown in light grey in both models, while the visible area of the disk (not shielded) to the clump is recolored to dark grey. In Patch Model (a), the radius ($r_p$) of the patch which is not drawn in scale is linearly proportional to the height ($H$) of the clump from the AD surface. In Wall Model (b), the radius ($r_w$) of the cylindrical shell is linearly proportional to the height ($h_w$) of the wall while $\beta=\tan(\theta)$ is constant. The visible area of the disk to the clump varies due to the changes in the size of patch and wall while the clump is moving. In order to keep the figures neat and clean, the conceptual location of $\mathrm{BLR}_{\mathrm{in}}$, and also the torus [$\mathrm{BLR}_{\mathrm{out}}$ to $R_{\mathrm{out}}$] are not drawn.}
\label{fig:ShModel}
\end{figure}

In this paper, for the first time we calculate the realistic 3-D radiation pressure of AD in which the radiation coming from other radii is also contributing. The real dust opacities are also taken into account, and the inner boundary condition of S-S AD is included. The general form of the 3-D equation of motion,
considering the forces acting on the clump to be the gravity of central blackhole and radiative force of AD, is thus given as below
\begin{equation}
\frac{d^2 \textbf{r}}{dt^2} =
- \frac{G M_{\mathrm{BH}}\ \textbf{r}}{r^3}
+ \int_{\lambda_i}^{\lambda_f}
\int_{\mathrm{visible\ area}} f\ (\textbf{r}, \psi, M_{\mathrm{BH}}, \dot{m},
\kappa_{\mathrm{ext}}(\lambda), T_{\mathrm{sub}}, \rho, \varphi, \lambda, C)\
d\textbf{a}\ d\lambda
\label{eq:motion}
\end{equation}
where $G$ is the gravitational constant, $M_{\mathrm{BH}}$ the blackhole mass, $\psi$ the dust-to-gas mass ratio, $\kappa_{\mathrm{ext}}$ the dust opacity as a function of wavelength, $\dot{m}$ the dimensionless accretion rate, $T_{\mathrm{sub}}$ the dust sublimation temperature, and $\textbf{r} = R \hat{R} + H \hat{z}$ is the vector showing the position of the clump in cylindrical coordinates in which the blackhole is at the origin as shown in the figure \ref{fig:ShModel}(a). The AD for simplicity is mapped onto the equatorial plane ($z=0$) from which the clump's height is measured. The parameters of $\rho$ and $\varphi$ show the coordinates of infinitesimal surface areas of AD in polar coordinates, and $C$ stands for some fixed factors and physical constants. The vector radiative force acting on a clump is calculated through a double integral evaluation over AD surface (\textit{visible area} due to shielding effect) and an integral over the effective range of wavelengths for the adopted dust model of the clump.

The effect of the dust evaporation is also included; assuming dust instantly emits the absorbed energy in the form of blackbody radiation, we numerically calculate the irradiated energy absorbed by dust $Q_{\mathrm{abs}}$ at each integration step. Once the $Q_{\mathrm{abs}}$ for a dusty clump reaching a certain altitude above AD exceeds the level of energy emission corresponding to sublimation temperature of dust $Q_{\mathrm{emit}}(T_{\mathrm{sub}})$, the clump loses its dust content. The radiation pressure force is then immediately switched off and the dustless cloud follows a ballistic motion, finally falling back onto the AD surface. The subsequent motion of ionized dustless gas driven by Thompson electron scattering is neglected.


\section{Shielding Effect}\label{sec:Shielding}

It is well-known from all radiatively-driven wind models that efficient alleviation of the material above the disk is not possible if the wind launching region is irradiated (see e.g. \citealp{2007ApJ...661..693P}). Some shielding is necessary, and it can be due to either inner failed Compton-driven wind or cold material forming at the edge between the inner hot ADAF (Advection Dominated Accretion Flow) and a cold disk. This shielding effect protects dusty material from too intense radiation from the central parts of AD, including X-rays. Therefore it would make the local radiation pressure from the disk to be effective in launching the material upward avoiding too early dust evaporation in the cloud. In order to mimic this effect we model it in the following two configurations:

\begin{itemize}

  \item Patch Model
  
  In this model it is assumed that  the  radiation  only comes from a small patch of the disk as shown in figure \ref{fig:ShModel}(a). The radius of patch is proportional to the height of the clump from the disk surface. Mathematically speaking, this dynamically variable radius can be expressed as a function of clump's height as $r_{p} = \alpha H $ where $\alpha$ is a free constant parameter of the model. The location of patch passively moves with the clump and increases/decreases in size as the cloud goes up/comes down.
  
  \item Wall Model
  
  In this model as shown in figure \ref{fig:ShModel}(b), as long as the clump is at rest on the AD surface, the radiation due to inner region of AD extending from inner radius of AD ($R_{\mathrm{in}}$) up to half of the radial position of the clump ($R/2$) is shielded. The model can be illustrated as a circular wall barrier (cylindrical shell) surrounding the inner region of the AD whose radius and height are $r_w$ and $h_w = \beta\ r_w$, respectively; where the model constant parameter $\beta=\tan(\theta)$. The range of variation for $r_w$ is $\mathrm{min}(r_w)= R_{\mathrm{in}}$ and $\mathrm{max}(r_w)= R_{\mathrm{out}}/2$. Moreover, until the clump to be launched, the wall also obscures the radiation of almost half of the AD located behind the wall (viewed from the clump). As the clump flies up, due to an anti-correlation between the wall's radius and the clump's height, the wall shifts inwards while $\beta$ is constant. As a result the clump sees a larger portion of AD irradiating. The moments when the clump is flying in an altitude $H > \beta R $ above AD, it can be irradiated by the whole AD as if there is no shielding effect.
   
\end{itemize}

Obviously, the proposed models of shielding are geometrically different. The results from these models can provide us with useful information about how shielding effect really looks like, and how effective it is in the physics of BLR and its dynamics. We incorporate these shielding models into our computations by imposing proper dynamic upper-lower integral limits while evaluating the radiative force produced by AD.

\section{Numerical Setup}\label{sec:Setup}

In order to perform the simulation, we consider the following setup. The blackhole mass is set to be of $10^8M_{\odot}$ \citep{2018ApJ...866..115P} which is almost the mean value of the blackhole mass in quasar catalog of \cite{2011ApJS..194...45S}.
The underlying disk is the S-S disk model which radiates locally as a blackbody. We consider a range of AD radii from $R_{\mathrm{in}} = 6 R_g$ to $R_{\mathrm{out}} = 10^6 R_g$ where $R_g$ is the gravitational radius of the blackhole defined as $R_g = G M_{BH} /c^2$.
The adopted dimensionless accretion rates ($\dot{m} = \dot{M} / \dot{M}_{\mathrm{Edd}}$ where $\dot{M}_{\mathrm{Edd}}$ is the Eddington accretion rate) are 0.01, 0.1, and 1.
Actual values of wavelength-dependent dust opacity are incorporated into our computation. These values are computed using MCDRT \citep{1997A&A...317..859S} and KOSMA--$\tau$ PDR \citep{2013A&A...549A..85R} codes which give the mean extinction cross-section per dust mass (equivalent to opacity) for different dust models with a given distribution of dust grain size and different materials.
We employ the MRN dust model \citep{1977ApJ...217..425M} which consists of silicate and graphite and has an identical grain size distribution for both dust populations. Silicate features are seen in the AGN spectra (e.g. \citealp{2007ApJ...666..806N}); graphite is more problematic, we would rather expect amorphous grains at the basis of the UV spectra  (see e.g. \citealp{2004MNRAS.348L..54C}), but we use MRN for simplicity.
Moreover, having assumed the gas and dust within the clump being strongly coupled we set the parameter $\psi$ equal to 0.005
which is the mean value for Milky Way \citep{1977ApJ...217..425M}. However, an estimate mean ratio of 0.008 is also adopted by others for a sample of AGNs \citep{2018ApJ...854..158S}.
It is also assumed that the dust content of clumps sublimates at a temperature of 1500 K which represents approximately the mean value for the different grain species and sizes (see \citealp{2018MNRAS.474.1970B}). The clumps in our model are launched with an azimuthal Keplerian velocity and zero vertical velocity from the surface of AD which has a thickness, $D(R)$.
The latter, which is function of radius, was computed separately for different initial values of AD size, accretion rate, black hole mass, and viscosity based on Rosseland mean opacity \citep{2016ApJ...832...15C, 1999MNRAS.305..481R} without the effect of self-gravity of AD itself being included. This implicitly also means that the height of the clump ($H$) never vanishes since our mathematically thin disk is at $z=0$. The Runge-Kutta method is then applied in order to integrate the equations of motion.

\section{BLR shape in FRADO}\label{sec:BLRsize}

In order to find the distribution of material above AD we need to determine the radial extension and height of BLR as follows.

The radial extension of BLR is by definition the range confined within $\mathrm{BLR}_{\mathrm{in}}$ to $\mathrm{BLR}_{\mathrm{out}}$. Theoretically, $\mathrm{BLR}_{\mathrm{in}}$ is the onset of BLR where the dust is instantly sublimated upon departure from AD. The outer radius of BLR, i.e. $\mathrm{BLR}_{\mathrm{out}}$, is where the dust survives irradiation by the whole disk even in spherically symmetric approximation of the radiation field. We determine this outer radius by finding the position where the sublimation height for a given radius is equal to the radius itself. Obviously, the shielding effect does not play a role at this large height.
In order to find these two radial ends of BLR, for each different initial conditions we introduce a dense grid of ($R$, $H$) pairs ranging as $R: [R_{\mathrm{in}}$ to $R_{\mathrm{out}}]$, and $H: [D (R)$ to $R_{\mathrm{out}}]$. We then numerically compute $Q_{\mathrm{abs}}$ at each grid point. The geometrical location where $Q_{\mathrm{abs}} = Q_{\mathrm{emit}}(T_{\mathrm{sub}})$ is a curved line called \textit{sublimation location}, $S(R)$, dividing the region above AD into two parts, above which dust can not survive irradiation.
The crossing radius of $S(R)$ and $D(R)$ yields the $\mathrm{BLR}_{\mathrm{in}}$; and $\mathrm{BLR}_{\mathrm{out}}$ can be found by setting $S(R)=R$.

As for the geometrical height of BLR, we observe the trajectory of clumps being launched from a set of logarithmically spaced initial launching radii ($R_{\mathrm{init}}$) ranging from $\mathrm{BLR}_{\mathrm{in}}$ to $\mathrm{BLR}_{\mathrm{out}}$. Neglecting the relaxation due to probable encounters between clumps and loss/gain of momentum, we find the maximum heights ($H_{\mathrm{peak}}$) the clump can attain per each initial launching radius within the set. However, in order to stress on the importance of the opening angle of the BLR clouds distribution, introducing the variable ($P = H_{\mathrm{peak}} / R_{\mathrm{peak}} $) where $R_{\mathrm{peak}}$ is the maximum height's corresponding radius, we find the maximum of $P$, i.e. $P_{\mathrm{peak}}$.

Therefore, the shape of BLR is specified by [$\mathrm{BLR}_{\mathrm{in}} - \mathrm{BLR}_{\mathrm{out}}$], and [$D (R) - H_{\mathrm{peak}} (R)$]. 
The $R(P_{\mathrm{peak}})$ and $H(P_{\mathrm{peak}})$ corresponding to the unique value of $P_{\mathrm{peak}}$ represents the effective location of BLR clouds. This is based on the assumption that the the parameter $R(P_{\mathrm{peak}})$ marks the region effectively exposed to the irradiation by the central parts of AD.

\begin{table}
    \caption{Output data for the size of BLR for blackhole mass of $10^8 M_{\odot}$. Values of $L_{5100}$ are in ergs/s. Other data are in light-days except $\dot{m}$ and $P_{\mathrm{peak}}$ being dimensionless. Patch model and Wall model are specified by their characteristic parameters of $\alpha$ and $\beta$, respectively.}
    \centering
\begin{tabular}{|l|crrrr|ccc|}
\hline
Shielding Model & & & & & & \multicolumn{3}{c}{($i = 39.2$)} \\
  & $\dot{m}$ & $\mathrm{BLR}_{\mathrm{in}}$ &
$\mathrm{BLR}_{\mathrm{out}}$ & $P_{\mathrm{peak}}$ &
$R(P_{\mathrm{peak}})$ & $\log(L_{5100})$ & $\log(\tau_{\mathrm{\ in}})$ & $\log(\tau_{\mathrm{\ peak}})$ \\
\hline

        & $0.01$ & $4.95$ & $112.53$ & $0.027$ & $4.96$ & $43.81$ & $0.9067>$ & $0.9027$ \\
        $\alpha = 1.0$
        & $0.10$ & $9.45$ & $356.71$ & $0.105$ & $9.64$ & $44.52$ & $1.1867>$ & $1.1761$ \\
        & $1.00$ & $18.00$ & $1126.75$ & $0.358$ & $27.42$ & $45.21$ & $1.4639<$ & $1.5894$ \\
\hline

        & $0.01$ & $5.19$ & $112.53$ & $0.031$ & $5.21$ & $43.81$ & $0.9273>$ & $0.9233$ \\
        $\alpha = 1.5$ 
        & $0.10$ & $9.92$ & $356.71$ & $0.117$ & $10.25$ & $44.52$ & $1.2078>$ & $1.2005$ \\
        & $1.00$ & $18.89$ & $1126.75$ & $0.351$ & $46.97$ & $45.21$ & $1.4853<$ & $1.8241$ \\
\hline

        & $0.01$ & $5.31$ & $112.53$ & $0.032$ & $5.32$ & $43.81$ & $0.9372>$ & $0.9321$ \\
        $\alpha = 2.0$
        & $0.10$ & $10.13$ & $356.71$ & $0.122$ & $10.61$ & $44.52$ & $1.2169>$ & $1.2146$ \\
        & $1.00$ & $19.29$ & $1126.75$ & $0.346$ & $^{(1)}31.72$ & $45.21$ & $1.4944<$ & $1.6543$ \\
\hline

        $\alpha = 2.0$ ($10^7 M_{\odot}$)
        & $0.10$ & $2.06$ & $-$ & $0.045$ & $2.07$ & $43.21$ & $0.5256>$ & $0.5196$ \\
\hline        
        
        $\alpha = 2.0$ ($10^9 M_{\odot}$)
        & $0.10$ & $49.76$ & $-$ & $0.270$ & $75.25$ & $45.81$ & $1.9067<$ & $2.0228$ \\
\hline

        & $0.01$ & $5.49$ & $112.53$ & $0.034$ & $5.63$ & $43.81$ & $0.9517<$ & $0.9563$ \\
        $\beta = 0.3$
        & $0.10$ & $10.48$ & $356.71$ & $0.115$ & $12.71$ & $44.52$ & $1.2319<$ & $1.2943$ \\
        & $1.00$ & $20.03$ & $1126.75$ & $0.245$ & $^{(1)}39.84$ & $45.21$ & $1.5111<$ & $1.7682$ \\
\hline

        & $0.01$ & $5.49$ & $112.53$ & $0.033$ & $5.64$ & $43.81$ & $0.9517<$ & $0.9573$ \\
        $\beta = 0.2$
        & $0.10$ & $10.48$ & $356.71$ & $0.106$ & $13.40$ & $44.52$ & $1.2319<$ & $1.3190$ \\
        & $1.00$ & $20.03$ & $1126.75$ & $0.199$ & $^{(1)}33.80$ & $45.21$ & $1.5111<$ & $1.7043$ \\
\hline

        & $0.01$ & $5.49$ & $112.53$ & $0.033$ & $5.70$ & $43.81$ & $0.9517<$ & $0.9619$ \\
        $\beta = 0.1$
        & $0.10$ & $10.48$ & $356.71$ & $0.079$ & $^{(1)}14.58$ & $44.52$ & $1.2319<$ & $1.3607$ \\
        & $1.00$ & $20.03$ & $1126.75$ & $0.144$ & $^{(1)}28.02$ & $45.21$ & $1.5111<$ & $1.6324$ \\
\hline

        & $0.01$ & $9.72$ & $112.53$ & $0.007$ & $16.04$ & $43.81$ & $1.2000<$ & $1.4165$ \\
        No Shielding
        & $0.10$ & $37.90$ & $356.71$ & $...$ & $^{(2)}...$ & $...$ & $...$ & $...$ \\
        & $1.00$ & $155.27$ & $1126.75$ & $...$ & $^{(2)}...$ & $...$ & $...$ & $...$ \\
\hline

        & $0.01$ & $7.63$ & $ 72.08$ & $-$ & $-$ & $-$ & $-$ & $-$ \\
        $^{(3)}$Analytical
        & $0.10$ & $16.44$ & $227.93$ & $-$ & $-$ & $-$ & $-$ & $-$ \\
        \quad FRADO
        & $1.00$ & $35.42$ & $720.79$ & $-$ & $-$ & $-$ & $-$ & $-$ \\
\hline
\multicolumn{8}{l}{\small $(1)$ Reaching the peak with dust-content sublimated: dust-less failed wind}\\
\multicolumn{8}{l}{\small $(2)$ It never forms a failed wind neither dust-less nor dusty. All clumps just escape to torus}\\
\multicolumn{8}{l}{\small $(3)$ This row is just added for the purpose of comparison.} 

\end{tabular}
    \label{simulationtable}
\end{table}

\section{Results \& Discussion}\label{sec:Results}

\begin{figure}
\begin{center}
\includegraphics[scale=0.52]{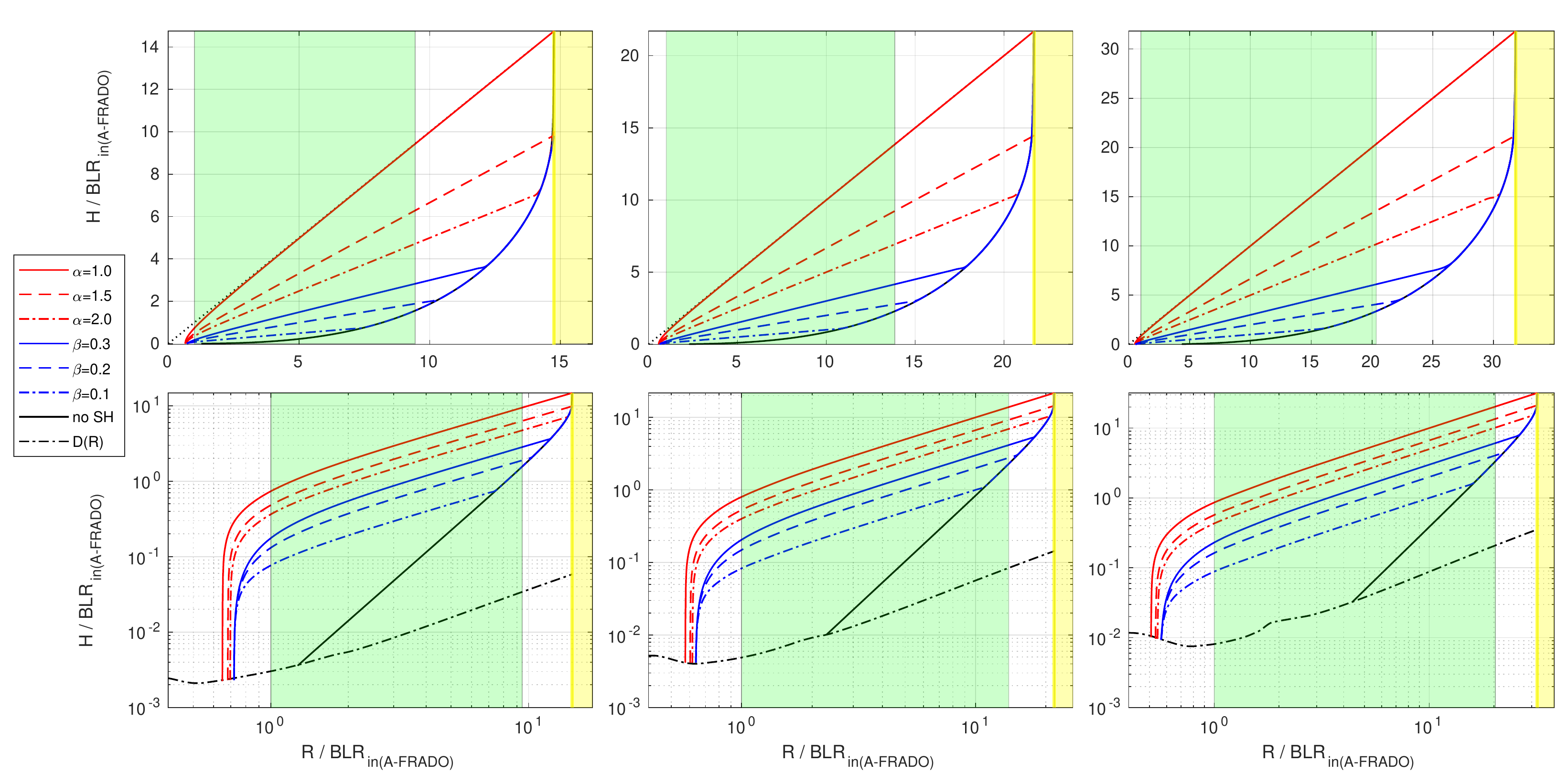}
\end{center}
\caption{Representation of the sublimation location above AD equatorial plane for Eddington rates of $0.01$(left panels), $0.1$ (middle panels), $1$ (right panels) in normal (upper panels) and logarithmic (lower panels) scales. Axes are normalized to $\mathrm{BLR}_{\mathrm{in}}$ from analytical FRADO model for the corresponding Eddington rate to be easily comparable. The area shaded in light-green covers the radial extension of BLR predicted by analytical FRADO model. Yellowish strip shows the inner part of torus. The onset of BLR and the thickness of AD are clearly visible in logarithmic panels, while in the linear ones the BLR radial extension for different accretion rates is easily comparable.}
\label{fig:Sublim}
\end{figure}

\begin{figure}
\begin{center}
\includegraphics[scale=0.92]{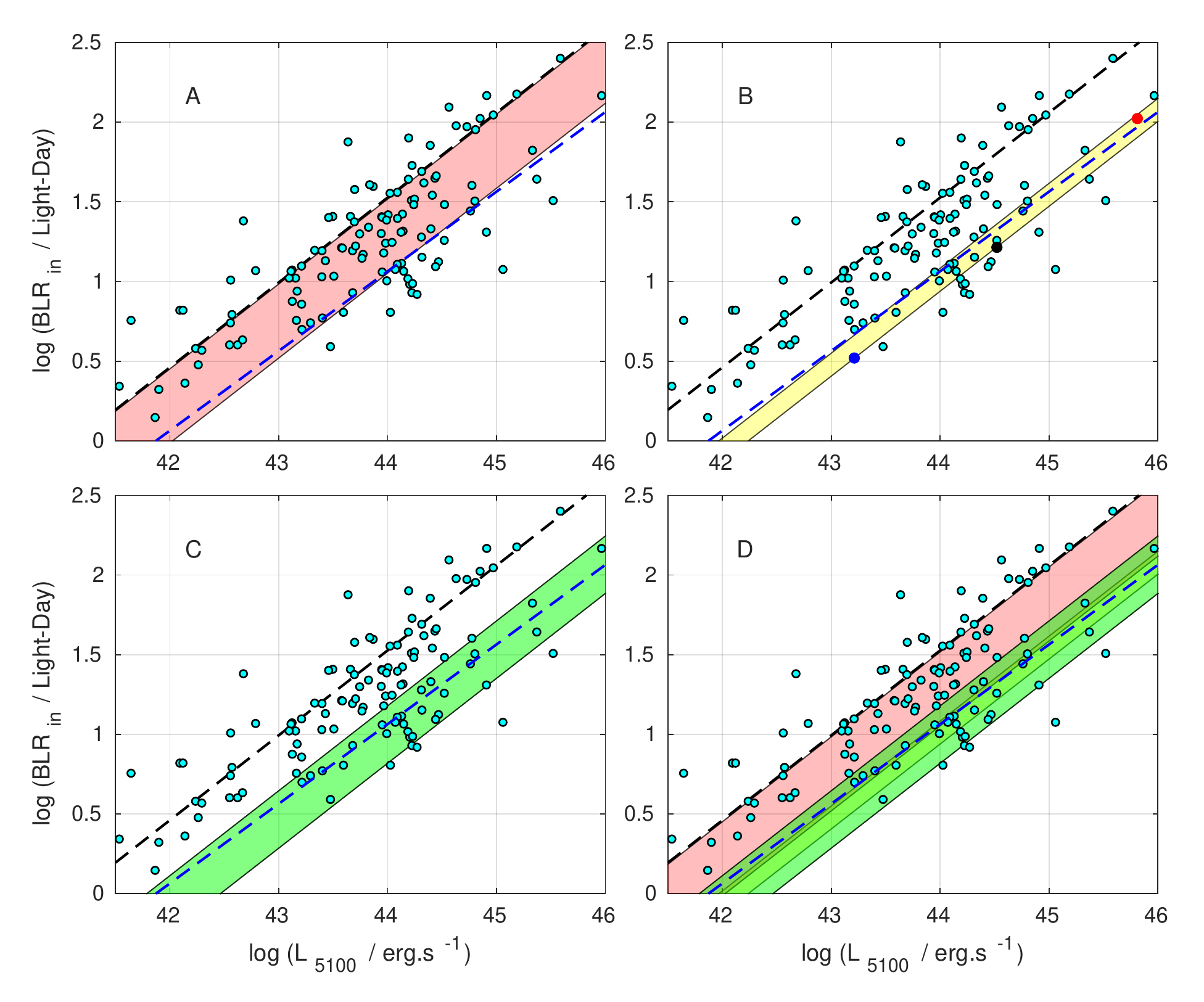}
\end{center}
\caption{
RL relation (considering the \citet{2013ApJ...767..149B} relation's slope) based on all reliable time-delays obtained from numerical FRADO  for all shielding cases and Eddington rates of 0.01 (A), 0.1 (B), 1 (C), blackhole mass of $10^8$, and $i=39.2$. Panel D shows the overlap of all Eddington rates. The shaded areas cover the range between the minimum and maximum reliable time-delays obtained per each Eddington rate.
Points in cyan show a sample of observational data from \citet{2019ApJ...883..170M}. The blue dashed line represents the RL relation predicted by analytical FRADO. The black dashed line shows the RL relation calibrated for sources with low Eddington ratios by \citet{2013ApJ...767..149B}.
Three specific solutions for of $\alpha=2.0$ and accretion rate of $0.1$ are shown as points in panel B with the blue, black, and red points corresponding to blackhole mass of $10^7$, $10^8$, $10^9 M_{\odot}$, respectively.
}
\label{fig:RLrelation}
\end{figure}

Having performed numerical computations for different initial conditions and for both models of shielding, we have obtained values for the size of BLR as collected in the table \ref{simulationtable}. Numerically computed values of the monochromatic luminosity of AD at $5100\ \textup{\AA}$ in erg/s depend only on the accretion rate and blackhole mass; the effects of the shielding and AD self-irradiation are not included \citep{2003A&A...412..317C}. Other columns of data listed in the table \ref{simulationtable} are expressed in light-days, except $\dot{m}$ and $P_{\mathrm{peak}}$ being dimensionless. The viewing angle for calculating the time-delay $\tau$ (equivalent to BLR radius assuming the light speed $c$ propagation of the signals) is set to be $i=39.2$ \citep{2010ApJ...714..561L}.

The sublimation surface, $S(R)$, obtained for all initial conditions and parameters of shielding models in table \ref{simulationtable} is shown in figure \ref{fig:Sublim}. Upper and lower panels show the same plots but in linear and log scale, respectively; in the log plots the onset of BLR and disk thickness is clearly visible while the linear plots better illustrate the radial extension of BLR for different values of the accretion rate. Obviously, the Patch model provides a stronger shielding than Wall model, and the table is also sorted down based on the strength of shielding. 
However, it seems that the presented shielding models can behave similarly if certain values of $\alpha$ and $\beta$ are chosen.
For example, based on the figure \ref{fig:Sublim} one can expect the two shielding models to show very similar patterns for $\alpha = 1/\beta$.
$S(R)$ profiles in figure \ref{fig:Sublim} for all values of $\alpha.$ and $\beta$ show a convergence to the case of no-shielding at $\mathrm{BLR}_{\mathrm{out}}$. This is related to the definition of the $\mathrm{BLR}_{\mathrm{out}}$, with the requested condition set at large height of the cloud above the disk where the shielding becomes unimportant. In addition the wall model interestingly shows a privilege of unique prediction of $\mathrm{BLR}_{\mathrm{in}}$ for all values of $\beta$ that implies the Wall shielding does not work efficiently when the cloud height is small as it happens close to $\mathrm{BLR}_{\mathrm{in}}$. However, testing with other values of $\beta$ is required to firmly determine the model properties. 

The two considered shielding models have significantly different geometrical properties, as shown in figure \ref{fig:ShModel}. In the Wall model a cloud is exposed to the radiation from a broad range of radii of AD while in the Patch model it is affected only by a local radiation. This may affect the cloud dynamics, and likely will be reflected in the emission-lines shape which is beyond the scope of the current work. However, further studies including the cases with $\alpha = 1/\beta$ are required to highlight the differences between these shielding models properly, and to enable us to make a choice between these two.

As can be seen in figure \ref{fig:Sublim}, the ratio of ($\mathrm{BLR}_{\mathrm{out}} / \mathrm{BLR}_{\mathrm{in}}$) increases with Eddington ratio for both analytical and numerical FRADO. However, the BLR size in numerical models with shielding is radially more extended than that of analytical ones which implies longer time-delays from the entire BLR.

According to the table \ref{simulationtable}, in case of no-shielding for accretion rates of $0.1$ and $1$, all launched clumps only escape to the torus and do not form a failed wind. These cases were omitted due to being out of interest in the current preliminary study. 
There are cases for accretion rates of $0.01$ and $0.1$ in Patch model in the table \ref{simulationtable} with $R_{\mathrm{peak}} \sim \mathrm{BLR}_{\mathrm{in}}$ yielding time-delays for $\mathrm{BLR}_{\mathrm{in}}$, $\log(\tau_{\mathrm{\ in}})$, being larger than that of $R_{\mathrm{peak}}$, $\log(\tau_{\mathrm{\ peak}})$. They imply the presence of a very sharp cut-off in dynamics of material past the $\mathrm{BLR}_{\mathrm{in}}$. These cases can not be considered as reliable. It is while for the accretion rate of $1$, all shielding cases give a larger value for $\log(\tau_{\mathrm{\ peak}})$. This means that for high accretion rate the presence of shielding is necessary. In other words, the higher the accretion rate is, the stronger shielding is required. A denser set of bins in the range of values of $\alpha$ and $\beta$ in future studies will be more indicative.

As explained in section \ref{sec:BLRsize}, since larger values of opening angle are of our interest, we chose the model showing the maximum $P_{\mathrm{peak}}$ among the obtained values for all shielding models per the arbitrary accretion rate of $0.1$ to perform a test with other blackhole masses. This maximum $P_{\mathrm{peak}}$ corresponds to $\alpha=2.0$, however it does not necessarily mean the preference of Patch model over Wall model, or appropriateness of this $\alpha$. We then repeated the simulation for this model for two other blackhole masses of $10^7$, and $10^9 M_{\odot}$. Interestingly, while the Patch model with the adopted values of $\alpha$, was too strong as mentioned above to be considered as reliable for accretion rates of $0.01$ and $0.1$, the consistent time-delay obtained for the blackhole mass of $10^9 M_{\odot}$ and accretion rate of $0.1$ with $\alpha=2.0$ indicates the dependence of shielding on blackhole mass as well; i.e. the higher the blackhole mass is, the stronger shielding is required.

As for plotting the RL relation, since we only have the data for the blackhole mass of $10^8M_{\odot}$ (except the case of $\alpha=2.0$), we take use of the slope of \citet{2013ApJ...767..149B} relation.
The RL relation based on reliable results for all adopted values of accretion rates are then shown in figure \ref{fig:RLrelation}, and compared to observational data and analytical FRADO model (see also \citealt{2019arXiv191200278N} for RL relation based on $\tau_{\mathrm{\ in}}$). Time-delays are computed assuming the effective reflection of reprocessed radiation of continuum happens for the clouds located in farther side of AD, as in \citet{2011A&A...525L...8C}, while the clouds in the closer side are almost self-obscured. As the whole material in the range between the onset of BLR and the maximum height can be irradiated by the continuum, the area between minimum/maximum time-delays corresponding to the minimum  $\mathrm{BLR}_{\mathrm{in}}$/maximum $R(P_{\mathrm{peak}})$ obtained out of all reliable data in the table \ref{simulationtable} for a given accretion rate are shown as shaded patterns in the figure \ref{fig:RLrelation}.
While the analytical FRADO only gives one line, the realistic 3-D approach covers almost the whole sample of observational data.
The points corresponding to three blackhole masses for the test case of $\alpha=2.0$ with accretion rate of $0.1$ are also depicted in the figure \ref{fig:RLrelation}(B) which are located within the yellow strip, and follow the RL trend apparently. However testing with a wider-denser set of blackhole masses is also required in future studies to provide us with a full insight into the models.

\section{Conclusions}\label{sec:Conclusion}

In this preliminary work we theoretically studied the dynamics of matter in the BLR based on a realistic 3-D approach to FRADO model. We introduced two geometrical configurations for shielding effect, and based on the dynamics, we tried to predict the BLR size (radial extension and geometrical height) and its dependence on different Eddington ratios [and blackhole mass] to compare with observational data. The main findings can be concluded as below

\begin{itemize}
    \item FRADO model can account for the range of measured time-delays of H$\beta$ line in AGNs. The model is expected to work for all Low Ionization Lines (LIL) like Hbeta and Mg-II \citep{2019ApJ...880...46C, 2019AN....340..577Z} but not for High Ionization Lines (HIL) which form closer in and come from line-driven wind part.
    \item The shielding must be included in the model, although low Eddington sources do not require it.
    \item The role of shielding must increase with the Eddington ratio of the source, as well as with blackhole mass.
    \item BLR size is predicted to be radially more extended compared to analytical FRADO which implies longer time-delays from the entire BLR.
    \item Future modelling of the emission line shapes based on the cloud dynamics will bring further constraints to the shielding geometry.
\end{itemize}

\section*{Funding}
The project was partially supported by National Science Centre, Poland, grant No. 2017/26/A/ST9/00756 (Maestro 9), and by the Ministry of Science and Higher Education (MNiSW) grant DIR/WK/2018/12.

\section*{Acknowledgments}
We are grateful to Mary Loli Martinez-Aldama for providing us with the observational data used in Figure~3. Authors would like to thank the anonymous referees for their fruitful comments on the manuscript that improved the quality of the paper.

\bibliographystyle{frontiersinSCNS_ENG_HUMS}
\bibliography{frontiers}

\begin{thebibliography}{52}
\providecommand{\natexlab}[1]{#1}
\expandafter\ifx\csname urlstyle\endcsname\relax
  \providecommand{\doi}[1]{doi:\discretionary{}{}{}#1}\else
  \providecommand{\doi}{doi:\discretionary{}{}{}\begingroup
  \urlstyle{rm}\Url}\fi
\providecommand{\selectlanguage}[1]{\relax}
\providecommand{\bibAnnoteFile}[1]{%
  \IfFileExists{#1}{\begin{quotation}\noindent\textsc{Key:} #1\\
  \textsc{Annotation:}\ \input{#1}\end{quotation}}{}}
\providecommand{\bibAnnote}[2]{%
  \begin{quotation}\noindent\textsc{Key:} #1\\
  \textsc{Annotation:}\ #2\end{quotation}}

\bibitem[{{Baskin} and {Laor}(2018)}]{2018MNRAS.474.1970B}
{Baskin}, A. and {Laor}, A. (2018).
\newblock {Dust inflated accretion disc as the origin of the broad line region
  in active galactic nuclei}.
\newblock \emph{MNRAS} 474, 1970--1994.
\newblock \doi{10.1093/mnras/stx2850}
\bibAnnoteFile{2018MNRAS.474.1970B}

\bibitem[{{Begelman} et~al.(1983){Begelman}, {McKee}, and
  {Shields}}]{1983ApJ...271...70B}
{Begelman}, M.~C., {McKee}, C.~F., and {Shields}, G.~A. (1983).
\newblock {Compton heated winds and coronae above accretion disks. I.
  Dynamics.}
\newblock \emph{ApJ} 271, 70--88.
\newblock \doi{10.1086/161178}
\bibAnnoteFile{1983ApJ...271...70B}

\bibitem[{{Bentz} et~al.(2013){Bentz}, {Denney}, {Grier}, {Barth}, {Peterson},
  {Vestergaard} et~al.}]{2013ApJ...767..149B}
{Bentz}, M.~C., {Denney}, K.~D., {Grier}, C.~J., {Barth}, A.~J., {Peterson},
  B.~M., {Vestergaard}, M., et~al. (2013).
\newblock {The Low-luminosity End of the Radius-Luminosity Relationship for
  Active Galactic Nuclei}.
\newblock \emph{ApJ} 767, 149.
\newblock \doi{10.1088/0004-637X/767/2/149}
\bibAnnoteFile{2013ApJ...767..149B}

\bibitem[{{Bentz} et~al.(2009){Bentz}, {Walsh}, {Barth}, {Baliber}, {Bennert},
  {Canalizo} et~al.}]{2009ApJ...705..199B}
{Bentz}, M.~C., {Walsh}, J.~L., {Barth}, A.~J., {Baliber}, N., {Bennert},
  V.~N., {Canalizo}, G., et~al. (2009).
\newblock {The Lick AGN Monitoring Project: Broad-line Region Radii and Black
  Hole Masses from Reverberation Mapping of H{\ensuremath{\beta}}}.
\newblock \emph{ApJ} 705, 199--217.
\newblock \doi{10.1088/0004-637X/705/1/199}
\bibAnnoteFile{2009ApJ...705..199B}

\bibitem[{{Blandford} and {Begelman}(1999)}]{1999MNRAS.303L...1B}
{Blandford}, R.~D. and {Begelman}, M.~C. (1999).
\newblock {On the fate of gas accreting at a low rate on to a black hole}.
\newblock \emph{MNRAS} 303, L1--L5.
\newblock \doi{10.1046/j.1365-8711.1999.02358.x}
\bibAnnoteFile{1999MNRAS.303L...1B}

\bibitem[{{Blandford} and {Payne}(1982)}]{1982MNRAS.199..883B}
{Blandford}, R.~D. and {Payne}, D.~G. (1982).
\newblock {Hydromagnetic flows from accretion disks and the production of radio
  jets.}
\newblock \emph{MNRAS} 199, 883--903.
\newblock \doi{10.1093/mnras/199.4.883}
\bibAnnoteFile{1982MNRAS.199..883B}

\bibitem[{{Collin} and {Zahn}(1999)}]{1999A&A...344..433C}
{Collin}, S. and {Zahn}, J.-P. (1999).
\newblock {Star formation and evolution in accretion disks around massive black
  holes.}
\newblock \emph{A\&A} 344, 433--449
\bibAnnoteFile{1999A&A...344..433C}

\bibitem[{{Collin} and {Zahn}(2008)}]{2008A&A...477..419C}
{Collin}, S. and {Zahn}, J.~P. (2008).
\newblock {Star formation in accretion discs: from the Galactic center to
  active galactic nuclei}.
\newblock \emph{A\&A} 477, 419--435.
\newblock \doi{10.1051/0004-6361:20078191}
\bibAnnoteFile{2008A&A...477..419C}

\bibitem[{{Czerny} et~al.(2016){Czerny}, {Du}, {Wang}, and
  {Karas}}]{2016ApJ...832...15C}
{Czerny}, B., {Du}, P., {Wang}, J.-M., and {Karas}, V. (2016).
\newblock {A Test of the Formation Mechanism of the Broad Line Region in Active
  Galactic Nuclei}.
\newblock \emph{ApJ} 832, 15.
\newblock \doi{10.3847/0004-637X/832/1/15}
\bibAnnoteFile{2016ApJ...832...15C}

\bibitem[{{Czerny} and {Hryniewicz}(2011)}]{2011A&A...525L...8C}
{Czerny}, B. and {Hryniewicz}, K. (2011).
\newblock {The origin of the broad line region in active galactic nuclei}.
\newblock \emph{A\&A} 525, L8.
\newblock \doi{10.1051/0004-6361/201016025}
\bibAnnoteFile{2011A&A...525L...8C}

\bibitem[{{Czerny} et~al.(2004){Czerny}, {Li}, {Loska}, and
  {Szczerba}}]{2004MNRAS.348L..54C}
{Czerny}, B., {Li}, J., {Loska}, Z., and {Szczerba}, R. (2004).
\newblock {Extinction due to amorphous carbon grains in red quasars from the
  Sloan Digital Sky Survey}.
\newblock \emph{MNRAS} 348, L54--L57.
\newblock \doi{10.1111/j.1365-2966.2004.07590.x}
\bibAnnoteFile{2004MNRAS.348L..54C}

\bibitem[{{Czerny} et~al.(2017){Czerny}, {Li}, {Hryniewicz}, {Panda}, {Wildy},
  {Sniegowska} et~al.}]{2017ApJ...846..154C}
{Czerny}, B., {Li}, Y.-R., {Hryniewicz}, K., {Panda}, S., {Wildy}, C.,
  {Sniegowska}, M., et~al. (2017).
\newblock {Failed Radiatively Accelerated Dusty Outflow Model of the Broad Line
  Region in Active Galactic Nuclei. I. Analytical Solution}.
\newblock \emph{ApJ} 846, 154.
\newblock \doi{10.3847/1538-4357/aa8810}
\bibAnnoteFile{2017ApJ...846..154C}

\bibitem[{{Czerny} et~al.(2015){Czerny}, {Modzelewska}, {Petrogalli}, {Pych},
  {Adhikari}, {{\.Z}ycki} et~al.}]{2015AdSpR..55.1806C}
{Czerny}, B., {Modzelewska}, J., {Petrogalli}, F., {Pych}, W., {Adhikari},
  T.~P., {{\.Z}ycki}, P.~T., et~al. (2015).
\newblock {The dust origin of the Broad Line Region and the model consequences
  for AGN unification scheme}.
\newblock \emph{Advances in Space Research} 55, 1806--1815.
\newblock \doi{10.1016/j.asr.2015.01.004}
\bibAnnoteFile{2015AdSpR..55.1806C}

\bibitem[{{Czerny} et~al.(2003){Czerny}, {Niko{\l}ajuk},
  {R{\'o}{\.z}a{\'n}ska}, {Dumont}, {Loska}, and {Zycki}}]{2003A&A...412..317C}
{Czerny}, B., {Niko{\l}ajuk}, M., {R{\'o}{\.z}a{\'n}ska}, A., {Dumont}, A.~M.,
  {Loska}, Z., and {Zycki}, P.~T. (2003).
\newblock {Universal spectral shape of high accretion rate AGN}.
\newblock \emph{A\&A} 412, 317--329.
\newblock \doi{10.1051/0004-6361:20031441}
\bibAnnoteFile{2003A&A...412..317C}

\bibitem[{{Czerny} et~al.(2019){Czerny}, {Olejak}, {Ra{\l}owski},
  {Koz{\l}owski}, {Loli Martinez Aldama}, {Zajacek}
  et~al.}]{2019ApJ...880...46C}
{Czerny}, B., {Olejak}, A., {Ra{\l}owski}, M., {Koz{\l}owski}, S., {Loli
  Martinez Aldama}, M., {Zajacek}, M., et~al. (2019).
\newblock {Time Delay Measurement of Mg II Line in CTS C30.10 with SALT}.
\newblock \emph{ApJ} 880, 46.
\newblock \doi{10.3847/1538-4357/ab2913}
\bibAnnoteFile{2019ApJ...880...46C}

\bibitem[{{Czerny} and {King}(1989)}]{1989MNRAS.236..843C}
{Czerny}, M. and {King}, A.~R. (1989).
\newblock {Accretion disc winds and coronae}.
\newblock \emph{MNRAS} 236, 843--850.
\newblock \doi{10.1093/mnras/236.4.843}
\bibAnnoteFile{1989MNRAS.236..843C}

\bibitem[{{Dong} et~al.(2008){Dong}, {Wang}, {Wang}, {Yuan}, {Zhou}, {Dai}
  et~al.}]{2008MNRAS.383..581D}
{Dong}, X., {Wang}, T., {Wang}, J., {Yuan}, W., {Zhou}, H., {Dai}, H., et~al.
  (2008).
\newblock {Broad-line Balmer decrements in blue active galactic nuclei}.
\newblock \emph{MNRAS} 383, 581--592.
\newblock \doi{10.1111/j.1365-2966.2007.12560.x}
\bibAnnoteFile{2008MNRAS.383..581D}

\bibitem[{{Du} et~al.(2015){Du}, {Hu}, {Lu}, {Huang}, {Cheng}, {Qiu}
  et~al.}]{2015ApJ...806...22D}
{Du}, P., {Hu}, C., {Lu}, K.-X., {Huang}, Y.-K., {Cheng}, C., {Qiu}, J., et~al.
  (2015).
\newblock {Supermassive Black Holes with High Accretion Rates in Active
  Galactic Nuclei. IV. H{\ensuremath{\beta}} Time Lags and Implications for
  Super-Eddington Accretion}.
\newblock \emph{ApJ} 806, 22.
\newblock \doi{10.1088/0004-637X/806/1/22}
\bibAnnoteFile{2015ApJ...806...22D}

\bibitem[{{Du} et~al.(2016){Du}, {Lu}, {Zhang}, {Huang}, {Wang}, {Hu}
  et~al.}]{2016ApJ...825..126D}
{Du}, P., {Lu}, K.-X., {Zhang}, Z.-X., {Huang}, Y.-K., {Wang}, K., {Hu}, C.,
  et~al. (2016).
\newblock {Supermassive Black Holes with High Accretion Rates in Active
  Galactic Nuclei. V. A New Size-Luminosity Scaling Relation for the Broad-line
  Region}.
\newblock \emph{ApJ} 825, 126.
\newblock \doi{10.3847/0004-637X/825/2/126}
\bibAnnoteFile{2016ApJ...825..126D}

\bibitem[{{Du} et~al.(2018){Du}, {Zhang}, {Wang}, {Huang}, {Zhang}, {Lu}
  et~al.}]{2018ApJ...856....6D}
{Du}, P., {Zhang}, Z.-X., {Wang}, K., {Huang}, Y.-K., {Zhang}, Y., {Lu}, K.-X.,
  et~al. (2018).
\newblock {Supermassive Black Holes with High Accretion Rates in Active
  Galactic Nuclei. IX. 10 New Observations of Reverberation Mapping and
  Shortened H{\ensuremath{\beta}} Lags}.
\newblock \emph{ApJ} 856, 6.
\newblock \doi{10.3847/1538-4357/aaae6b}
\bibAnnoteFile{2018ApJ...856....6D}

\bibitem[{{Elitzur} et~al.(2014){Elitzur}, {Ho}, and
  {Trump}}]{2014MNRAS.438.3340E}
{Elitzur}, M., {Ho}, L.~C., and {Trump}, J.~R. (2014).
\newblock {Evolution of broad-line emission from active galactic nuclei}.
\newblock \emph{MNRAS} 438, 3340--3351.
\newblock \doi{10.1093/mnras/stt2445}
\bibAnnoteFile{2014MNRAS.438.3340E}

\bibitem[{{Emmering} et~al.(1992){Emmering}, {Blandford}, and
  {Shlosman}}]{1992ApJ...385..460E}
{Emmering}, R.~T., {Blandford}, R.~D., and {Shlosman}, I. (1992).
\newblock {Magnetic Acceleration of Broad Emission-Line Clouds in Active
  Galactic Nuclei}.
\newblock \emph{ApJ} 385, 460.
\newblock \doi{10.1086/170955}
\bibAnnoteFile{1992ApJ...385..460E}

\bibitem[{{Grier} et~al.(2017){Grier}, {Trump}, {Shen}, {Horne}, {Kinemuchi},
  {McGreer} et~al.}]{2017ApJ...851...21G}
{Grier}, C.~J., {Trump}, J.~R., {Shen}, Y., {Horne}, K., {Kinemuchi}, K.,
  {McGreer}, I.~D., et~al. (2017).
\newblock {The Sloan Digital Sky Survey Reverberation Mapping Project:
  H{\ensuremath{\alpha}} and H{\ensuremath{\beta}} Reverberation Measurements
  from First-year Spectroscopy and Photometry}.
\newblock \emph{ApJ} 851, 21.
\newblock \doi{10.3847/1538-4357/aa98dc}
\bibAnnoteFile{2017ApJ...851...21G}

\bibitem[{{Haas} et~al.(2011){Haas}, {Chini}, {Ramolla}, {Pozo Nu{\~n}ez},
  {Westhues}, {Watermann} et~al.}]{2011A&A...535A..73H}
{Haas}, M., {Chini}, R., {Ramolla}, M., {Pozo Nu{\~n}ez}, F., {Westhues}, C.,
  {Watermann}, R., et~al. (2011).
\newblock {Photometric AGN reverberation mapping - an efficient tool for BLR
  sizes, black hole masses, and host-subtracted AGN luminosities}.
\newblock \emph{A\&A} 535, A73.
\newblock \doi{10.1051/0004-6361/201117325}
\bibAnnoteFile{2011A&A...535A..73H}

\bibitem[{{Hu} et~al.(2008){Hu}, {Wang}, {Ho}, {Chen}, {Zhang}, {Bian}
  et~al.}]{2008ApJ...687...78H}
{Hu}, C., {Wang}, J.-M., {Ho}, L.~C., {Chen}, Y.-M., {Zhang}, H.-T., {Bian},
  W.-H., et~al. (2008).
\newblock {A Systematic Analysis of Fe II Emission in Quasars: Evidence for
  Inflow to the Central Black Hole}.
\newblock \emph{ApJ} 687, 78--96.
\newblock \doi{10.1086/591838}
\bibAnnoteFile{2008ApJ...687...78H}

\bibitem[{{Kaspi} et~al.(2000){Kaspi}, {Smith}, {Netzer}, {Maoz}, {Jannuzi},
  and {Giveon}}]{2000ApJ...533..631K}
{Kaspi}, S., {Smith}, P.~S., {Netzer}, H., {Maoz}, D., {Jannuzi}, B.~T., and
  {Giveon}, U. (2000).
\newblock {Reverberation Measurements for 17 Quasars and the
  Size-Mass-Luminosity Relations in Active Galactic Nuclei}.
\newblock \emph{ApJ} 533, 631--649.
\newblock \doi{10.1086/308704}
\bibAnnoteFile{2000ApJ...533..631K}

\bibitem[{{Lawrence} and {Elvis}(2010)}]{2010ApJ...714..561L}
{Lawrence}, A. and {Elvis}, M. (2010).
\newblock {Misaligned Disks as Obscurers in Active Galaxies}.
\newblock \emph{ApJ} 714, 561--570.
\newblock \doi{10.1088/0004-637X/714/1/561}
\bibAnnoteFile{2010ApJ...714..561L}

\bibitem[{{Mart{\'\i}nez-Aldama} et~al.(2019){Mart{\'\i}nez-Aldama}, {Czerny},
  {Kawka}, {Karas}, {Panda}, {Zaja{\v{c}}ek} et~al.}]{2019ApJ...883..170M}
{Mart{\'\i}nez-Aldama}, M.~L., {Czerny}, B., {Kawka}, D., {Karas}, V., {Panda},
  S., {Zaja{\v{c}}ek}, M., et~al. (2019).
\newblock {Can Reverberation-measured Quasars Be Used for Cosmology?}
\newblock \emph{ApJ} 883, 170.
\newblock \doi{10.3847/1538-4357/ab3728}
\bibAnnoteFile{2019ApJ...883..170M}

\bibitem[{{Mathis} et~al.(1977){Mathis}, {Rumpl}, and
  {Nordsieck}}]{1977ApJ...217..425M}
{Mathis}, J.~S., {Rumpl}, W., and {Nordsieck}, K.~H. (1977).
\newblock {The size distribution of interstellar grains.}
\newblock \emph{ApJ} 217, 425--433.
\newblock \doi{10.1086/155591}
\bibAnnoteFile{1977ApJ...217..425M}

\bibitem[{{Murray} et~al.(1995){Murray}, {Chiang}, {Grossman}, and
  {Voit}}]{1995ApJ...451..498M}
{Murray}, N., {Chiang}, J., {Grossman}, S.~A., and {Voit}, G.~M. (1995).
\newblock {Accretion Disk Winds from Active Galactic Nuclei}.
\newblock \emph{ApJ} 451, 498.
\newblock \doi{10.1086/176238}
\bibAnnoteFile{1995ApJ...451..498M}

\bibitem[{{Naddaf} et~al.(2019){Naddaf}, {Czerny}, and
  {Szczerba}}]{2019arXiv191200278N}
{Naddaf}, M.-H., {Czerny}, B., and {Szczerba}, R. (2019).
\newblock {R-L Relation in Realistic FRADO Model}.
\newblock \emph{arXiv e-prints} , arXiv:1912.00278
\bibAnnoteFile{2019arXiv191200278N}

\bibitem[{{Netzer}(2015)}]{2015ARA&A..53..365N}
{Netzer}, H. (2015).
\newblock {Revisiting the Unified Model of Active Galactic Nuclei}.
\newblock \emph{ARA\&A} 53, 365--408.
\newblock \doi{10.1146/annurev-astro-082214-122302}
\bibAnnoteFile{2015ARA&A..53..365N}

\bibitem[{{Netzer} et~al.(2007){Netzer}, {Lutz}, {Schweitzer}, {Contursi},
  {Sturm}, {Tacconi} et~al.}]{2007ApJ...666..806N}
{Netzer}, H., {Lutz}, D., {Schweitzer}, M., {Contursi}, A., {Sturm}, E.,
  {Tacconi}, L.~J., et~al. (2007).
\newblock {Spitzer Quasar and ULIRG Evolution Study (QUEST). II. The Spectral
  Energy Distributions of Palomar-Green Quasars}.
\newblock \emph{ApJ} 666, 806--816.
\newblock \doi{10.1086/520716}
\bibAnnoteFile{2007ApJ...666..806N}

\bibitem[{{Panda} et~al.(2018){Panda}, {Czerny}, {Adhikari}, {Hryniewicz},
  {Wildy}, {Kuraszkiewicz} et~al.}]{2018ApJ...866..115P}
{Panda}, S., {Czerny}, B., {Adhikari}, T.~P., {Hryniewicz}, K., {Wildy}, C.,
  {Kuraszkiewicz}, J., et~al. (2018).
\newblock {Modeling of the Quasar Main Sequence in the Optical Plane}.
\newblock \emph{ApJ} 866, 115.
\newblock \doi{10.3847/1538-4357/aae209}
\bibAnnoteFile{2018ApJ...866..115P}

\bibitem[{{Panda} et~al.(2019){Panda}, {Mart{\'\i}nez-Aldama}, and
  {Zaja{\v{c}}ek}}]{2019FrASS...6...75P}
{Panda}, S., {Mart{\'\i}nez-Aldama}, M.~L., and {Zaja{\v{c}}ek}, M. (2019).
\newblock {Current and future applications of Reverberation-mapped quasars in
  Cosmology}.
\newblock \emph{Frontiers in Astronomy and Space Sciences} 6, 75.
\newblock \doi{10.3389/fspas.2019.00075}
\bibAnnoteFile{2019FrASS...6...75P}

\bibitem[{{Peterson} et~al.(2004){Peterson}, {Ferrarese}, {Gilbert}, {Kaspi},
  {Malkan}, {Maoz} et~al.}]{2004ApJ...613..682P}
{Peterson}, B.~M., {Ferrarese}, L., {Gilbert}, K.~M., {Kaspi}, S., {Malkan},
  M.~A., {Maoz}, D., et~al. (2004).
\newblock {Central Masses and Broad-Line Region Sizes of Active Galactic
  Nuclei. II. A Homogeneous Analysis of a Large Reverberation-Mapping
  Database}.
\newblock \emph{ApJ} 613, 682--699.
\newblock \doi{10.1086/423269}
\bibAnnoteFile{2004ApJ...613..682P}

\bibitem[{{Peterson} et~al.(1998){Peterson}, {Wanders}, {Horne}, {Collier},
  {Alexander}, {Kaspi} et~al.}]{1998PASP..110..660P}
{Peterson}, B.~M., {Wanders}, I., {Horne}, K., {Collier}, S., {Alexander}, T.,
  {Kaspi}, S., et~al. (1998).
\newblock {On Uncertainties in Cross-Correlation Lags and the Reality of
  Wavelength-dependent Continuum Lags in Active Galactic Nuclei}.
\newblock \emph{PASA} 110, 660--670.
\newblock \doi{10.1086/316177}
\bibAnnoteFile{1998PASP..110..660P}

\bibitem[{{Proga}(2007)}]{2007ApJ...661..693P}
{Proga}, D. (2007).
\newblock {Dynamics of Accretion Flows Irradiated by a Quasar}.
\newblock \emph{ApJ} 661, 693--702.
\newblock \doi{10.1086/515389}
\bibAnnoteFile{2007ApJ...661..693P}

\bibitem[{{Rakshit} et~al.(2019){Rakshit}, {Woo}, {Gallo}, {Hodges-Kluck},
  {Shin}, {Jeon} et~al.}]{2019ApJ...886...93R}
{Rakshit}, S., {Woo}, J.-H., {Gallo}, E., {Hodges-Kluck}, E., {Shin}, J.,
  {Jeon}, Y., et~al. (2019).
\newblock {The Seoul National University AGN Monitoring Project. II. BLR Size
  and Black Hole Mass of Two AGNs}.
\newblock \emph{ApJ} 886, 93.
\newblock \doi{10.3847/1538-4357/ab49fd}
\bibAnnoteFile{2019ApJ...886...93R}

\bibitem[{{Risaliti} and {Elvis}(2010)}]{2010A&A...516A..89R}
{Risaliti}, G. and {Elvis}, M. (2010).
\newblock {A non-hydrodynamical model for acceleration of line-driven winds in
  active galactic nuclei}.
\newblock \emph{A\&A} 516, A89.
\newblock \doi{10.1051/0004-6361/200912579}
\bibAnnoteFile{2010A&A...516A..89R}

\bibitem[{{R{\"o}llig} et~al.(2013){R{\"o}llig}, {Szczerba}, {Ossenkopf}, and
  {Gl{\"u}ck}}]{2013A&A...549A..85R}
{R{\"o}llig}, M., {Szczerba}, R., {Ossenkopf}, V., and {Gl{\"u}ck}, C. (2013).
\newblock {Full SED fitting with the KOSMA-{\ensuremath{\tau}} PDR code. I.
  Dust modelling}.
\newblock \emph{A\&A} 549, A85.
\newblock \doi{10.1051/0004-6361/201118190}
\bibAnnoteFile{2013A&A...549A..85R}

\bibitem[{{R{\'o}{\.z}a{\'n}ska} et~al.(1999){R{\'o}{\.z}a{\'n}ska}, {Czerny},
  {{\.Z}ycki}, and {Pojma{\'n}ski}}]{1999MNRAS.305..481R}
{R{\'o}{\.z}a{\'n}ska}, A., {Czerny}, B., {{\.Z}ycki}, P.~T., and
  {Pojma{\'n}ski}, G. (1999).
\newblock {Vertical structure of accretion discs with hot coronae in active
  galactic nuclei}.
\newblock \emph{MNRAS} 305, 481--491.
\newblock \doi{10.1046/j.1365-8711.1999.02425.x}
\bibAnnoteFile{1999MNRAS.305..481R}

\bibitem[{{Shakura} and {Sunyaev}(1973)}]{1973A&A....24..337S}
{Shakura}, N.~I. and {Sunyaev}, R.~A. (1973).
\newblock {Black holes in binary systems. Observational appearance.}
\newblock \emph{A\&A} 500, 33--51
\bibAnnoteFile{1973A&A....24..337S}

\bibitem[{{Shangguan} et~al.(2018){Shangguan}, {Ho}, and
  {Xie}}]{2018ApJ...854..158S}
{Shangguan}, J., {Ho}, L.~C., and {Xie}, Y. (2018).
\newblock {On the Gas Content and Efficiency of AGN Feedback in Low-redshift
  Quasars}.
\newblock \emph{ApJ} 854, 158.
\newblock \doi{10.3847/1538-4357/aaa9be}
\bibAnnoteFile{2018ApJ...854..158S}

\bibitem[{{Shen} et~al.(2011){Shen}, {Richards}, {Strauss}, {Hall},
  {Schneider}, {Snedden} et~al.}]{2011ApJS..194...45S}
{Shen}, Y., {Richards}, G.~T., {Strauss}, M.~A., {Hall}, P.~B., {Schneider},
  D.~P., {Snedden}, S., et~al. (2011).
\newblock {A Catalog of Quasar Properties from Sloan Digital Sky Survey Data
  Release 7}.
\newblock \emph{ApJs} 194, 45.
\newblock \doi{10.1088/0067-0049/194/2/45}
\bibAnnoteFile{2011ApJS..194...45S}

\bibitem[{{Szczerba} et~al.(1997){Szczerba}, {Omont}, {Volk}, {Cox}, and
  {Kwok}}]{1997A&A...317..859S}
{Szczerba}, R., {Omont}, A., {Volk}, K., {Cox}, P., and {Kwok}, S. (1997).
\newblock {IRAS 22272+5435 - a source with 30 and 21{\ensuremath{\mu}}m
  features.}
\newblock \emph{A\&A} 317, 859--870
\bibAnnoteFile{1997A&A...317..859S}

\bibitem[{{Wang} et~al.(2012){Wang}, {Du}, {Baldwin}, {Ge}, {Hu}, and
  {Ferland}}]{2012ApJ...746..137W}
{Wang}, J.-M., {Du}, P., {Baldwin}, J.~A., {Ge}, J.-Q., {Hu}, C., and
  {Ferland}, G.~J. (2012).
\newblock {Star Formation in Self-gravitating Disks in Active Galactic Nuclei.
  II. Episodic Formation of Broad-line Regions}.
\newblock \emph{ApJ} 746, 137.
\newblock \doi{10.1088/0004-637X/746/2/137}
\bibAnnoteFile{2012ApJ...746..137W}

\bibitem[{{Wang} et~al.(2017){Wang}, {Du}, {Brotherton}, {Hu}, {Songsheng},
  {Li} et~al.}]{2017NatAs...1..775W}
{Wang}, J.-M., {Du}, P., {Brotherton}, M.~S., {Hu}, C., {Songsheng}, Y.-Y.,
  {Li}, Y.-R., et~al. (2017).
\newblock {Tidally disrupted dusty clumps as the origin of broad emission lines
  in active galactic nuclei}.
\newblock \emph{Nature Astronomy} 1, 775--783.
\newblock \doi{10.1038/s41550-017-0264-4}
\bibAnnoteFile{2017NatAs...1..775W}

\bibitem[{{Wang} et~al.(2011){Wang}, {Ge}, {Hu}, {Baldwin}, {Li}, {Ferland}
  et~al.}]{2011ApJ...739....3W}
{Wang}, J.-M., {Ge}, J.-Q., {Hu}, C., {Baldwin}, J.~A., {Li}, Y.-R., {Ferland},
  G.~J., et~al. (2011).
\newblock {Star Formation in Self-gravitating Disks in Active Galactic Nuclei.
  I. Metallicity Gradients in Broad-line Regions}.
\newblock \emph{ApJ} 739, 3.
\newblock \doi{10.1088/0004-637X/739/1/3}
\bibAnnoteFile{2011ApJ...739....3W}

\bibitem[{{Watson} et~al.(2011){Watson}, {Denney}, {Vestergaard}, and
  {Davis}}]{2011ApJ...740L..49W}
{Watson}, D., {Denney}, K.~D., {Vestergaard}, M., and {Davis}, T.~M. (2011).
\newblock {A New Cosmological Distance Measure Using Active Galactic Nuclei}.
\newblock \emph{ApJ} 740, L49.
\newblock \doi{10.1088/2041-8205/740/2/L49}
\bibAnnoteFile{2011ApJ...740L..49W}

\bibitem[{{Witt} et~al.(1997){Witt}, {Czerny}, and
  {Zycki}}]{1997MNRAS.286..848W}
{Witt}, H.~J., {Czerny}, B., and {Zycki}, P.~T. (1997).
\newblock {Accretion discs with accreting coronae in active galactic nuclei -
  II. The nuclear wind}.
\newblock \emph{MNRAS} 286, 848--864.
\newblock \doi{10.1093/mnras/286.4.848}
\bibAnnoteFile{1997MNRAS.286..848W}

\bibitem[{{Zaja{\v{c}}ek} et~al.(2019){Zaja{\v{c}}ek}, {Czerny},
  {Mart{\'\i}nez-Aldama}, and {Karas}}]{2019AN....340..577Z}
{Zaja{\v{c}}ek}, M., {Czerny}, B., {Mart{\'\i}nez-Aldama}, M.~L., and {Karas},
  V. (2019).
\newblock {Reverberation mapping of distant quasars: Time lag determination
  using different methods}.
\newblock \emph{Astronomische Nachrichten} 340, 577--585.
\newblock \doi{10.1002/asna.201913659}
\bibAnnoteFile{2019AN....340..577Z}

\end{thebibliography}

\end{document}